\documentclass[3p,times,twocolumn]{elsarticle}

\usepackage{ecrc}
\pdfoutput=1
\volume{00}

\firstpage{1}

\journalname{Nuclear Physics B Proceedings Supplement}

\runauth{}

\jid{nuphbp}
\jnltitlelogo{Nuclear Physics B Proceedings Supplement}

\usepackage{amssymb}
 \usepackage{lineno}

\usepackage[figuresright]{rotating}

\begin{document}
\begin{frontmatter}
\dochead{}

\title{$J/\psi, \Upsilon(1S)$ and $\chi_{b}(3p)$ Production Measurement with the ATLAS Detector}

\author{Rui Wang}
\address{Department of Physics and Astronomy, University of New Mexico\\ Albuquerque, NM, USA, 87131}
\author{On behalf of the ATLAS Collaboration.}

\begin{abstract}
The $J/\psi$  and $\Upsilon(1S)$ production cross-sections are measured in proton-proton collisions using the ATLAS detector at the LHC. Differential cross-sections are measured as a function of transverse momentum and rapidity. Results are compared to QCD predictions. A new $\chi_{b}$ state has been observed though radiative transitions to the $\Upsilon (1S)$ and $\Upsilon (2S)$ states.
\end{abstract}


\end{frontmatter}

% \linenumbers

\section{Introduction}
The production of heavy quarkonium at hadron colliders provides particular challenges and opportunities for insight into the theory of Quantum Chromodynamics (QCD) as its mechanisms of production operate at the boundary of the perturbative and non-perturbative regimes. Despite being among the most studied of the bound-quark systems, there is still no clear understanding of the mechanisms in the production of quarkonium states like the $J/\psi$ and $\Upsilon$ that can consistently explain both the production cross-section and spin-alignment measurements. Data obtained by the Large Hadron Collider (LHC) can help to test existing theoretical models of both quarkonium production and b-production in a new energy regime.\\
The most important elements of the ATLAS detector for B physics measurements are the Inner Detector (ID) tracker and the Muon Spectrometer, details can be found in \cite{detector}. Dedicated B physics triggers are based on both single muons and di-muons with different thresholds and mass ranges.\\
\\
\\
\\
\section{Measurement of the differential cross-sections for inclusive, prompt and non-prompt $J/\psi$ production}
The inclusive $J/\psi$  production cross-section is measured in the $J/\psi \rightarrow \mu^+ \mu^-$ decay channel as a function of both $J/\psi$ transverse momentum and rapidity using 2.3 pb$^{-1}$ of ATLAS 2010 7 TeV data \cite{jpsi}. \\
The ATLAS detector has a three-level trigger system: level 1 (L1), level 2 (L2) and the event filter (EF). In this measurement, the trigger relies on the Minimum Bias Trigger Scintillators (MBTS) and the muon trigger chambers. The MBTS trigger is configured to require two hits above threshold from either side of the detector at the L1 level. Then a dedicated muon trigger at the EF level is required to confirm the candidate events chosen for this measurement. This trigger, initiated by the MBTS L1 trigger and searches for the presence of at least one track in the entire MS, is referred to as the EF minimum bias trigger. As the instantaneous luminosity of the collider increases, the trigger requirement switches from the EF minimum bias trigger to the EF muon trigger which seeded by this L1 muon trigger. The EF level pT cut of the EF muon trigger is 4 GeV initially and grows to 6 GeV later.\\
The $J/\psi$ yields are determined by the number of $J/\psi$ candidates which are extracted from the observed di-muon pairs, with weights applied to unfold the acceptance, detector resolution, tracking, reconstruction and trigger efficiencies in each $p_{T}$ and rapidity bin. Studies are performed to assess all relevant sources of systematic uncertainty on the measurement of the $J/\psi$ inclusive production cross-section. Because the spin alignment of the $J/\psi$ is unknown at the LHC, an envelope of all possible spin alignment assumptions is taken and assigned as an additional systematic uncertainty.\\
\begin{figure}[!htb]
 \centering
  \includegraphics[width=8cm]{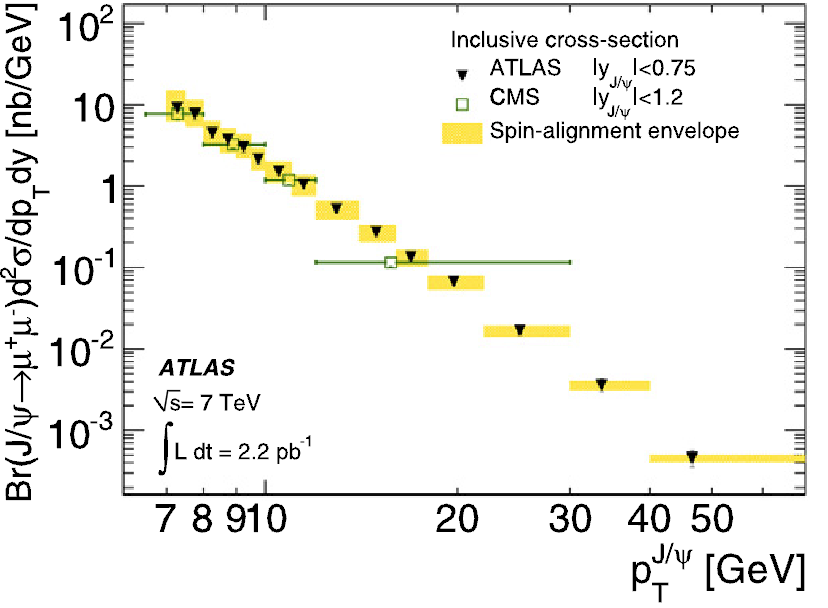}
  \caption{ The inclusive $J/\psi$ production cross-section as a function of $J/\psi$ transverse momentum in the $|y| < 0.75$ rapidity bin. The CMS result is included for comparison.}
  \label{fig:inclusivejpsi}
\end{figure}
Figure 1 shows the measurement of the differential cross-sections of inclusive $J/\psi$ production, with the CMS result \cite{CMS} included for comparison. The ATLAS and CMS results agree well in the overlapped regime. \\
\begin{figure}[!htb]
 \centering
  \includegraphics[width=8cm]{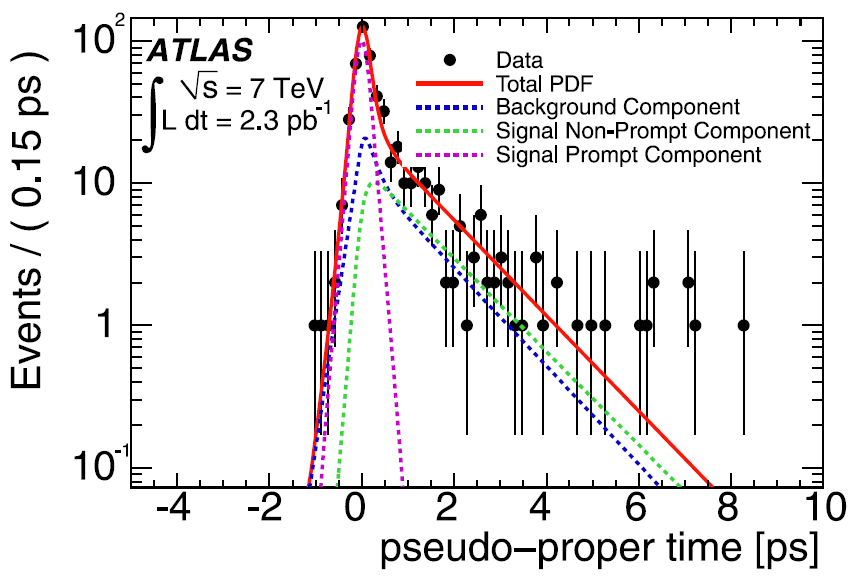}
  \caption{Pseudo-proper time distributions (top) of $J/\psi \rightarrow \mu^+ \mu^-$ candidates in the signal region, for example bin $9.5 < p_{T} < 10.0$ GeV and $|y| < 0.75$ region. The points with error bars are data. The solid line is the result of the maximum likelihood unbinned fit to all di-muon pairs in the 2.5 - 3.5 GeV mass region projected on the narrow mass window 2.9 - 3.3 GeV.}
  \label{fig:fit}
\end{figure}
It is possible to distinguished the prompt $J/\psi$'s produced directly from the p-p collision from the non-prompt $J/\psi$'s produced in B-hadron decays by the measurably displaced decay point due to the long lifetime of their B-hadron parent. A simultaneous unbinned maximum likelihood fit is applied to both invariant mass and lifetime (Figure 2), which allows us to determine the fraction of non-prompt $J/\psi$ as a function of $p_{T}$ and rapidity. The result for one of the rapidity slices is shown in Figure 3, in good agreement with the CMS result \cite{CMS}. This fraction is strongly dependent on $p_{T}^{J/\psi}$ with only weak dependence on $\eta^{J/\psi}$. The result has been compared to the CDF result \cite{CDF} indicating a limited dependence on collision energy. \\
\begin{figure}[!htb]
 \centering
  \includegraphics[width=8cm]{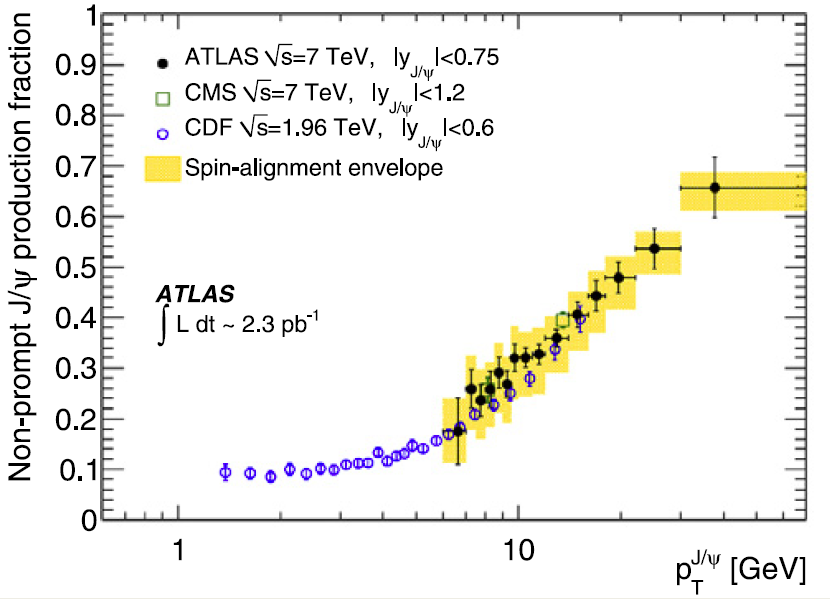}
  \caption{ The non-prompt $J/\psi$ to inclusive fraction as a function of $J/\psi$ transverse momentum. Overlaid is a band representing the variation of the result under various spin-alignment scenarios. The equivalent results from CMS and CDF are included.}
  \label{fig:fraction}
\end{figure}
By combining the results of the inclusive $J/\psi$ cross-section and non-prompt $J/\psi$ fraction in each $p_{T}$ and rapidity bin, the non-prompt and prompt $J/\psi$ differential cross-section can be extracted. The non-prompt $J/\psi$ results are in a good agreement with the Fixed Order Next-to-Leading-Log (FONLL) prediction \cite{FONLL1}\cite{FONLL2}. The prompt $J/\psi$ results are compared to the Colour Evaporation Model (CEM) \cite{CEM} and Colour Singlet Model(CSM) \cite{CSM1}\cite{CSM2}. The Colour Evaporation Model is showing significant disagreement in the extended $p_{T}$ range. The Colour Singlet Model with an NNLO calculation shows significant improvement in describing the $p_{T}$ dependence and normalization of prompt $J/\psi$ production over the NLO calculation but is still lower than the data. This is expected to be relatively significant for hidden charm production. Result in the regime $|y|<0.75$ is shown in Figure 4.\\
\begin{figure}[!htb]
  \centering
  \includegraphics[width=8cm]{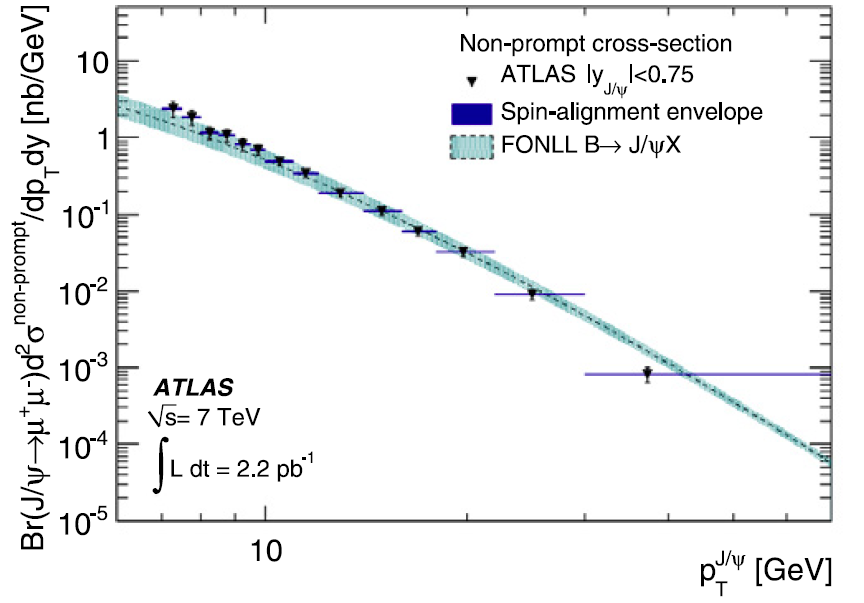}
  \includegraphics[width=8cm]{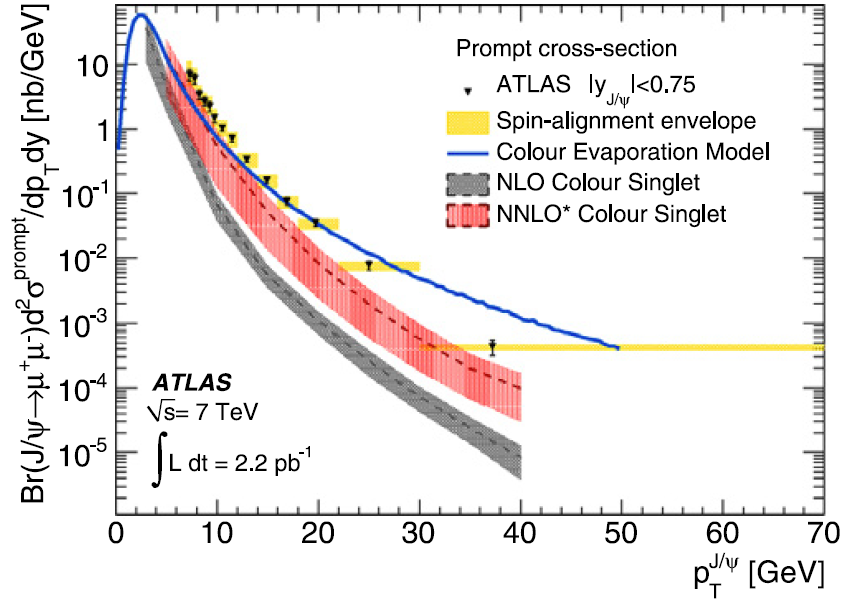}
  \caption{ Non-prompt (top) and prompt (bottom) $J/\psi$ production cross-sections as a function of $J/\psi$ transverse momentum with $|y|<0.75$. Non-prompt results are compared to FONLL predictions. Prompt results are compared to NLO and NNLO and Colour Evaporation Model predictions. Overlaid is a band representing the variation of the result under various spin-alignment assumptions on the non-prompt and prompt components. The central value assumes an isotropic polarization for both prompt and non-prompt production.}
  \label{fig:prompt}
\end{figure}
\section{Measurement of the $\Upsilon (1S)$ production cross-section}
\begin{figure}[!htb]
  \centering
  \includegraphics[width=8cm]{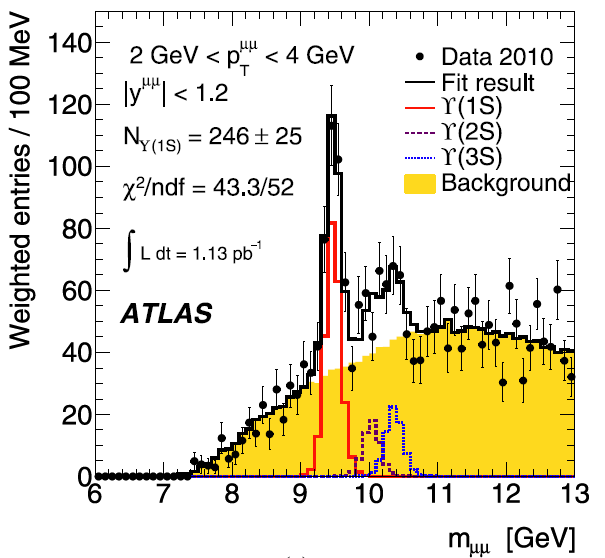}
  \caption{ Dimuon mass distributions in $2 < p_{T} < 4$ GeV and $1.2 < |y| < 2.4$ region. The data (filled circles) are shown together with the result of the unbinned maximum likelihood fit (histogram). The shaded histogram shows the background contribution, and the three other histograms show the contributions from the three $\Upsilon$ states. All histograms are normalised by the factor determined in the fit. In the individual plots, the fitted $N_{\Upsilon (1S)}$ yield with its statistical uncertainty, the $\chi^2$ and the number of degrees of freedom are also given. It should be noted that this is simply a binned graphical representation of the fit; the actual fit is unbinned and interpolates the template histograms to obtain the input probability density function.}
  \label{fig:upsilon}
\end{figure}
The $\Upsilon (1S)$  production cross-section is measured in the $\Upsilon (1S) \rightarrow \mu^+ \mu^-$ decay channel as a function of both $\Upsilon (1S)$ transverse momentum and rapidity using 1.13 pb$^{-1}$ of ATLAS 2010 7 TeV data \cite{upsilon}. Muons are selected via a single muon trigger with a threshold of 4 GeV and are required to have $p_{T} >$ 4 GeV and $|\eta|<2.5$ in order to remove the uncertainty associated with spin alignment in the cross-section measurement. The $\Upsilon (1S)$ yields are determined by correcting the number of reconstructed $\Upsilon (1S)$ candidates with weights which unfold the trigger and reconstruction efficiencies in each $p_{T}$ and rapidity bin. The
typical uncertainty is about 10 - 15$\%$ at low $p_{T}$ and 35$\%$ at high $p_{T}$ and is dominated by the statistical precision of the data.\\
The number of $\Upsilon (1S)$ events is determined from an unbinned maximum likelihood fit to the dimuon mass distributions in each bin. The three $\Upsilon$ signal templates are taken from the corresponding $\Upsilon$ MC samples. The background templates are constructed from data by pairing a muon with a track reconstructed in the ID of opposite electric charge (OS). This template (denoted as ¡°OS $\mu$ + track¡±) gives an adequate description of the background since its shape is primarily determined by the kinematic selection requirements. An example of the distributions and corresponding fit results is shown in Figure 5.\\
\begin{figure}[!htb]
  \centering
  \includegraphics[width=8cm]{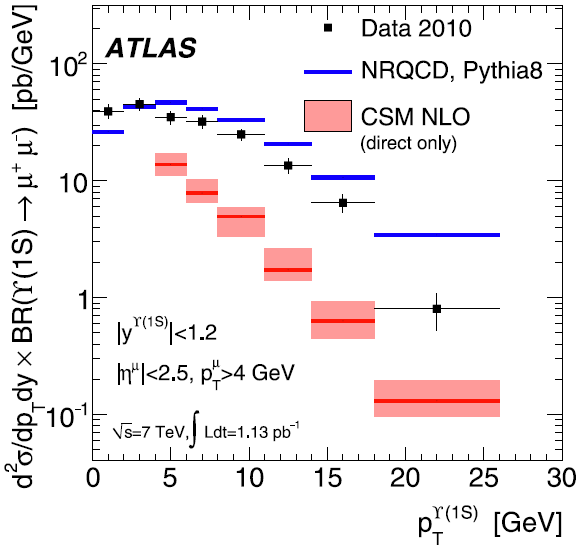}
  \includegraphics[width=8cm]{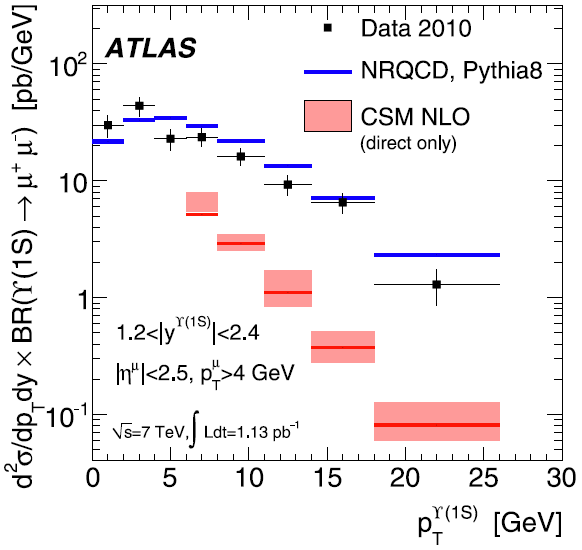}
  \caption{ The differential $\Upsilon (1S)$ cross-section for $|y^{\Upsilon (1S)} | < 1.2$ (top) and $1.2 < |y^{\Upsilon (1S)} | < 2.4$ (bottom) as a function of $p_{T}^{\Upsilon (1S)}$ for $p_{T}^{\mu} >$ 4 GeV and $|\eta^{¦Ì}| < 2.5$ on both muons. Also shown is the colour-singlet NLO (CSM) \cite{CSM} prediction using $m_{T} = \sqrt{4m_b^2+p_{T}^2}$ ($m_b$ = 4.75 GeV) for the renormalisation and factorization scales. The shaded area shows the change in the theoretical prediction when varying the renormalisation and factorization scales by a factor of two. The CSM NLO calculation accounts only for direct production of $\Upsilon (1S)$ mesons and not for any feed-down from excited states. The NRQCD prediction as implemented in Pythia8 \cite{pythia} is also shown for a particular choice of parameters.}
  \label{fig:upsilon}
\end{figure}
The results are compared to both NLO and NRQCD predictions (Figure 6). The data significantly exceed the NLO prediction but this may be explained by contributions from higher mass bound states and by the need for additional higher order corrections to $\Upsilon (1S)$ production. In contrast, the data are in reasonable agreement with the NRQCD prediction as implemented in Pythia8 but differences in the shape of the $p_{T}$ spectrum of about a factor of two are observed. These data will be useful to further understand the complex mechanisms that govern quarkonium production.\\
\section{Observation of a new $\chi_{b}$ state in radiative transitions to $\Upsilon(1S)$ and $\Upsilon(2S)$}
The $\chi_{b}(nP)$ quarkonium states are studied using a data sample corresponding to an integrated luminosity of 4.4 fb$^{-1}$ of ATLAS 2011 data \cite{chib}. Previous experiments have measured the $\chi_{b}(1P)$ and $\chi_{b}(2P)$ through decay modes of $\chi_{b}(nP)\rightarrow\Upsilon(1S)\gamma$ and $\chi_{b}(nP)\rightarrow\Upsilon(2S)\gamma$.\\
In this measurement, a set of muon triggers designed to select events containing muon pairs or single high transverse momentum muons was used to collect the data sample. For each muon candidate it must have a track, reconstructed in the MS, combined with a track reconstructed in the ID with $p_{T} >$ 4 GeV and pseudorapidity $|\eta| <$ 2.3. The di-muon selection requires a pair of oppositely charged muons, which are fitted to a common vertex. The di-muon candidate is also required to have $p_{T} >$ 12 GeV and $|\eta| <$ 2.0.\\
\begin{figure}[!htb]
  \centering
  \includegraphics[width=8cm]{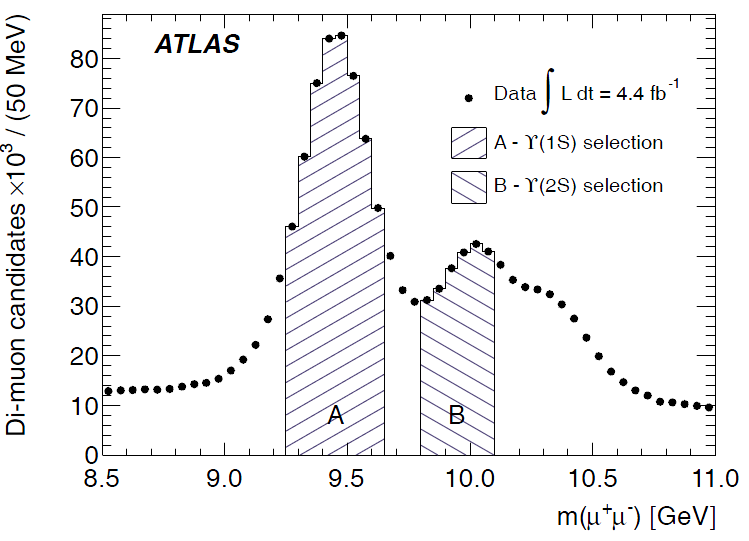}
  \caption{ The invariant mass of selected di-muon candidates. The shaded regions A and B show the selections for $\Upsilon(1S)$ and $\Upsilon(2S)$ candidates respectively.}
  \label{fig:upsilon}
\end{figure}
The $\Upsilon(1S)\rightarrow\mu\mu$ candidates with masses in the ranges $9.25 < m_{\mu\mu} < 9.65$ GeV and $\Upsilon(2S)\rightarrow\mu\mu$ candidates with masses in the ranges $9.80 < m_{\mu\mu} < 10.10$ GeV are selected(Figure 7). This asymmetric mass window for $\Upsilon(2S)$ candidates is chosen in order to reduce contamination from the $\Upsilon(3S)$ peak and continuum background contributions. A photon is combined with each $\Upsilon$ candidate. Both converted photons reconstructed by ID tracks from $e^+e^-$ pairs with a conversion vertex and unconverted photos reconstructed by electromagnetic calorimeter energy deposit are used. The converted photon candidates are required to be within $|\eta| <$ 2.30 while the unconverted photon candidates are required to be within $|\eta| <$ 2.37. Unconverted photons must also be outside the transition region between the barrel and the endcap calorimeters, 1.37 $< |\eta| <$ 1.52. Requirements of $p_{T}(\mu^+\mu^-) >$ 20 GeV and $p_{T}(\mu^+\mu^-) >$ 12 GeV are applied to $\Upsilon$ candidates with unconverted and converted photon candidates respectively. These thresholds are chosen in order to optimize signal significance in the $\chi_{b}(1P,2P)$ peaks.\\
\begin{figure}[!htb]
  \centering
  \includegraphics[width=8cm]{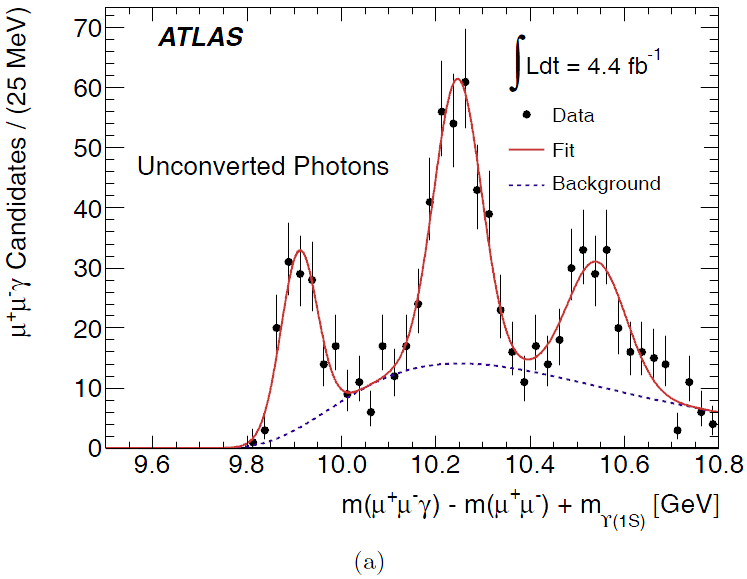}
  \includegraphics[width=8cm]{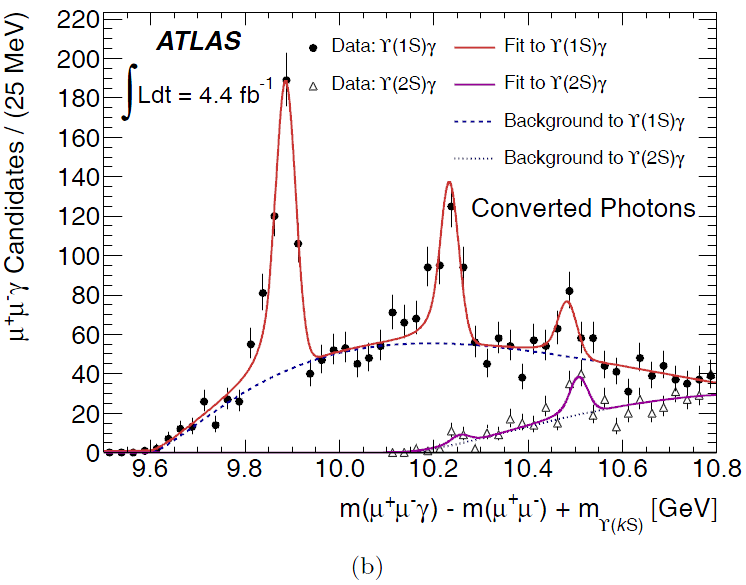}
  \caption{ (a) The mass distribution of $\chi_{b}\rightarrow\Upsilon(1S)\gamma$ candidates for unconverted photons reconstructed from energy deposits in the electromagnetic calorimeter. (b) The mass distributions of $\chi_{b}\rightarrow\Upsilon(kS)\gamma$ (k = 1, 2) candidates formed using photons which have converted and been reconstructed in the ID. Data are shown before the correction for the energy loss from the photon conversion electrons due to bremsstrahlung and other processes. The data for decays of $\chi_{b}\rightarrow\Upsilon(1S)\gamma$ and $\chi_{b}\rightarrow\Upsilon(2S)\gamma$ are plotted using circles and triangles respectively. Solid lines represent the total fit result for each mass window. The dashed lines represent the background components only.}
  \label{fig:upsilon}
\end{figure}
As shown in the mass difference $m(\mu^+\mu^-)-m(\mu^+\mu^-\gamma)$ distributions (Figure 8), in addition to the mass peaks corresponding to the decay modes $\chi_{b}(1P,2P)\rightarrow\Upsilon(1S)$, a new structure centered at mass $10.530 \pm 0.005 (stat.) \pm 0.009 (syst.)$ GeV is also observed. This is interpreted as the $\chi_{b}(3P)$ state.\\



%% References

\nocite{*}
\bibliographystyle{elsarticle-num}
\bibliography{beach2012}





% \end{thebibliography}

\end{document}